# BAYESIAN MODEL COMPARISON AND MODEL AVERAGING FOR SMALL-AREA ESTIMATION[1]

By Murray Aitkin, Charles C. Liu and Tom Chadwick

*University of Melbourne, University of Melbourne and Newcastle University*

This paper considers small-area estimation with lung cancer mortality data, and discusses the choice of upper-level model for the variation over areas. Inference about the random effects for the areas may depend strongly on the choice of this model, but this choice is not a straightforward matter. We give a general methodology for both evaluating the data evidence for different models and averaging over plausible models to give robust area effect distributions. We reanalyze the data of Tsutakawa [*Biometrics* **41** (1985) 69–79] on lung cancer mortality rates in Missouri cities, and show the differences in conclusions about the city rates from this methodology.

**1. The lung cancer data.** The data are male lung cancer mortality frequencies and population sizes for the period 1972–1981 in $N = 84$ Missouri cities (see Table 1). The variables, given in Tsutakawa and reproduced below, are the number $r$ of men aged 45–54 dying from lung cancer in each city over the period 1972–1981 and the city size $n$.

Most of the "cities" are small, though three are large. The mortality rates are poorly defined in small cities; four cities with populations less than 200 have no deaths at all, so the observed rate is zero. Our aim is to estimate the mortality rates in each city, using the information from other cities in the most appropriate way.

**2. Small-area estimation.** Variance component models are widely used in small-area estimation; the term *borrowing strength* is commonly used to

Received April 2008; revised August 2008.
[1]Supported in part by the Australian Research Council under Grant DP0559684 and in part by Federal funds from the U.S. Department of Education, National Center for Education Statistics under Contract RN95127001. The content of this publication does not necessarily reflect the views or policies of the U.S. Department of Education, National Center for Education Statistics, nor does mention of trade names, commercial products or organizations imply endorsement by the U.S. Government.

TABLE 1
*Male lung cancer mortality frequency and city size 1972–1981 in 84 Missouri cities*

| # | $n$ | $r$ | # | $n$ | $r$ | # | $n$ | $r$ | # | $n$ | $r$ |
|---|---|---|---|---|---|---|---|---|---|---|---|
| 1 | 1019 | 2 | 22 | 260 | 1 | 43 | 254 | 2 | 64 | 581 | 6 |
| 2 | 1512 | 8 | 23 | 371 | 2 | 44 | 28937 | 251 | 65 | 550 | 6 |
| 3 | 1424 | 8 | 24 | 232 | 1 | 45 | 445 | 4 | 66 | 431 | 5 |
| 4 | 54155 | 402 | 25 | 228 | 1 | 46 | 447 | 4 | 67 | 399 | 5 |
| 5 | 447 | 1 | 26 | 343 | 2 | 47 | 329 | 3 | 68 | 286 | 4 |
| 6 | 1907 | 12 | 27 | 454 | 3 | 48 | 206 | 2 | 69 | 592 | 7 |
| 7 | 1755 | 11 | 28 | 323 | 2 | 49 | 313 | 3 | 70 | 246 | 4 |
| 8 | 5756 | 42 | 29 | 311 | 2 | 50 | 314 | 3 | 71 | 547 | 7 |
| 9 | 509 | 2 | 30 | 784 | 6 | 51 | 314 | 3 | 72 | 438 | 6 |
| 10 | 350 | 1 | 31 | 426 | 3 | 52 | 202 | 2 | 73 | 202 | 4 |
| 11 | 473 | 2 | 32 | 184 | 1 | 53 | 198 | 2 | 74 | 790 | 10 |
| 12 | 329 | 1 | 33 | 181 | 1 | 54 | 183 | 2 | 75 | 648 | 9 |
| 13 | 7137 | 55 | 34 | 177 | 1 | 55 | 292 | 3 | 76 | 354 | 6 |
| 14 | 430 | 2 | 35 | 177 | 1 | 56 | 178 | 2 | 77 | 730 | 10 |
| 15 | 304 | 1 | 36 | 291 | 2 | 57 | 287 | 3 | 78 | 144 | 4 |
| 16 | 163 | 0 | 37 | 170 | 1 | 58 | 282 | 3 | 79 | 1093 | 14 |
| 17 | 163 | 0 | 38 | 158 | 1 | 59 | 164 | 2 | 80 | 384 | 7 |
| 18 | 159 | 0 | 39 | 274 | 2 | 60 | 164 | 2 | 81 | 278 | 6 |
| 19 | 281 | 1 | 40 | 150 | 1 | 61 | 1923 | 18 | 82 | 596 | 10 |
| 20 | 154 | 0 | 41 | 265 | 2 | 62 | 3672 | 34 | 83 | 1889 | 28 |
| 21 | 889 | 6 | 42 | 257 | 2 | 63 | 261 | 3 | 84 | 22514 | 334 |

describe the improvement in precision of estimation for the parameters of small areas by relating them through a two-level model. *Empirical Bayes* methods [Carlin and Louis (1996)] are widely used to represent the precision of the small-area parameters through their estimated posterior distributions, substituting the unknown variance component and other parameters by their maximum likelihood estimates. It has long been known [see, e.g., Tsutakawa (1985)] that this is not a satisfactory representation for the information about the small areas because the imprecision in the estimation of the variance component and other parameters is not allowed for. Fully Bayesian methods allow for this correctly, and will be discussed here.

The upper-level model for the among-area variation is routinely assumed to be normal on a suitable scale, though other choices, like the conjugate gamma distribution for Poisson rates, or the beta distribution for binomial proportions, are possible. With more than one possible model for rates, we need to compare the relative evidence for the competing models; if one is greatly superior to the others, we can discount the poorer models. A further issue is how to combine, or average over, the models if *several* are well supported by the data.

The comparison of upper-level models is frequently done using the Deviance Information Criterion DIC [Spiegelhalter et al. (2002)]. In this approach the *posterior distribution of the deviance* is computed for each model, and the mean (or sometimes the median) of this distribution is *penalized* by a function of the number of model parameters; an important issue in this approach is the need to specify the *focus* of the likelihood, in the sense of identifying the level in the model at which inference is to be performed.

This requirement was criticized by the paper discussants, and in more detail by Celeux et al. (2006); see also Trevisani and Gelfand (2003). In their response, Spiegelhalter et al. [(2002), page 634] said:

"Smith and others ask how the model focus should be chosen in practice. We argue that the focus is operationalized by the prediction problem of interest. For example, if the random effects $\theta$ in a hierarchical model relate to observation units such as schools or hospitals or geographical areas, where we might reasonably want to make future predictions for those same units, then taking $p(y|\theta)$ as the focus is sensible. The prediction problem is then to predict a new $Y_{i,\text{rep}}$ conditional on the posterior estimate of $\theta_i$ for that unit. However, if the random effects relate to individual people, say, then we are often interested in population-average inference rather than subject-specific inference, so we may want to predict responses for a new or 'typical' individual rather than an individual who is already in the data set. In this case, it is appropriate to integrate over the $\theta$s and to predict $Y_{\text{rep}}$ for a new individual conditional on [the random effect distribution parameters] $\psi$, leading to a model focused on $p(y|\psi)$. A crucial insight is that a predictive probability statement such as $p(Y_{\text{rep}}|y)$ is not uniquely defined without specifying the level of the hierarchy that is kept fixed in the prediction—this defines the focus of the model."

In the context of the lung cancer data, these two cases would correspond to inferential statements about a given city rate, and inference about an *average* or *typical* city rate.

In this paper we give a general methodology which treats these two cases in the same way, using the full posterior distribution of the Fisherian observed data likelihood (rather than a penalized version of its mean or median) and the standard marginal and conditional arguments for the individual city random effect distributions and the marginal rate distribution across cities.

We use this methodology to illustrate:

- the differences in conclusions which may arise from different upper-level models;
- the comparison of upper-level models;
- and the form of model averaging which protects against such differences,

for both individual city and "typical city" rates, with a re-analysis of the Missouri lung cancer data analyzed by Tsutakawa (1985).

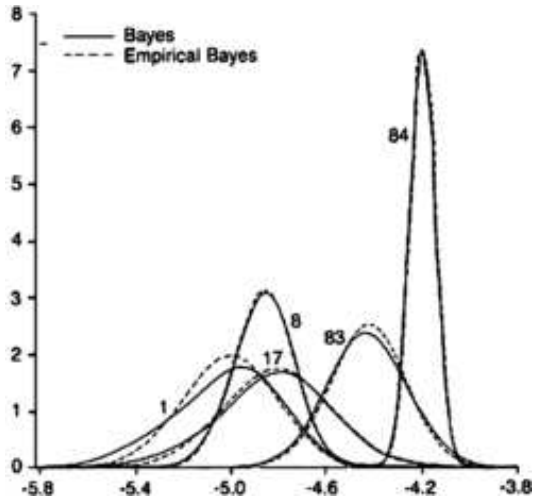

Fig. 1. *Posterior densities for five cities from Tsutakawa.*

**3. The Tsutakawa analysis.** Tsutakawa gave empirical Bayes and approximate fully Bayes analyses of these proportions, and showed a slight increase in dispersion of the fully Bayes posterior distributions for small cities. He used a nonconjugate normal model for the logit of the cancer rates which required several approximations for the computations.

His Figure 1, reproduced above, shows the empirical Bayes and full Bayes posteriors for the logits of the death rates for five of the cities, numbered 1, 8, 17, 83 and 84 in the data table.

The full Bayes posteriors are slightly more diffuse than the empirical Bayes posteriors, particularly for city 1 with its small population.

We note for subsequent comparison that Bayes inferences about the city rates may be based directly on the data from each city, without (apparently) *any* modeling: with a flat prior on the rate $p_i$ in the $i$th city, and the observed death count $r_i$ in the population of $n_i$, the $i$th city death rate has a posterior $Beta(r_i + 1, n_i - r_i + 1)$ distribution. (We do not use the Jeffreys or Haldane priors, as these give improper posteriors for the cities with zero cases.)

These distributions for the same five cities, shown on the logit scale in Figure 2, are quite different from the empirical or the full Bayes posteriors in Figure 1 for the small cities, though very similar for the larger cities. This is an example of "posterior shrinkage" from the normal model assumption for the distribution of the city random effects on the logit scale.

Other possible models need to be examined for the death rates and compared to the normal model, since posterior shrinkage may be quite different for different models [Aitkin (1999)]. We set out a general methodology for this purpose, following Dempster (1974, 1997), Aitkin (1997) and Aitkin, Boys and Chadwick (2005); see also Fox (2005).

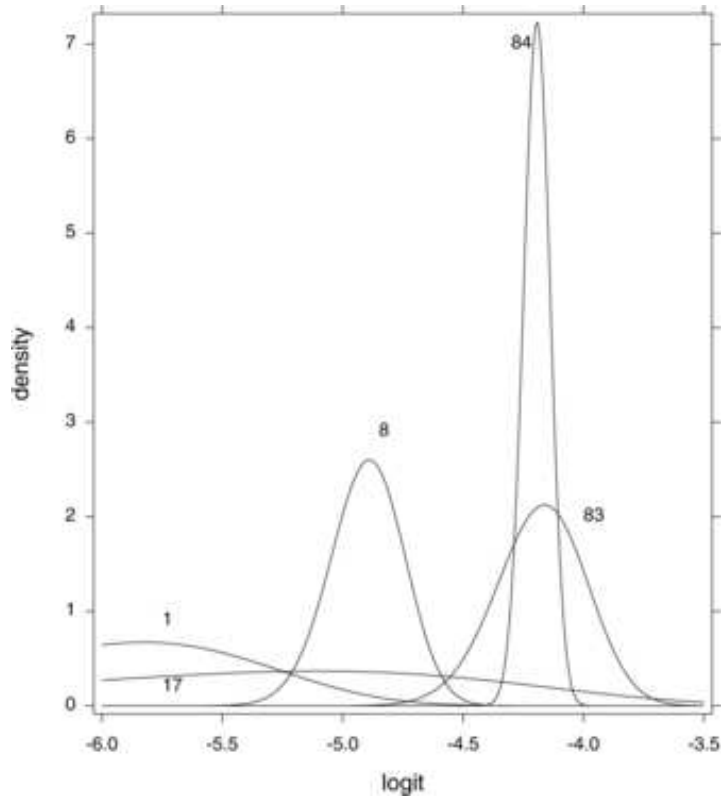

Fig. 2. *Posterior densities from each city data only.*

**4. Model comparison via posterior deviances.** We have competing models $M_j$ $(j = 1, \ldots, J)$ for observed data $\mathbf{y}$, with densities $f_j(y|\theta_j)$ under model $M_j$, and prior distributions $\pi_j(\theta_j)$. Given the data $\mathbf{y}$, we form the likelihoods $L_j(\theta_j|\mathbf{y})$, and update the priors to the posteriors $\pi_j(\theta_j|\mathbf{y})$. We make $T$ independent draws (both within and among models) $\theta_j^{[t]}$ from the posteriors, and substitute these into the likelihoods, to give $L_j(\theta_j^{[t]})$. These are $T$ independent draws from the posterior distributions of the likelihoods $L_j(\theta_j|\mathbf{y})$. To compare models $j$ and $k$, we compute the likelihood ratio values $L_{jk}^{[t]} = L_j(\theta_j^{[t]})/L_k(\theta_k^{[t]})$, which are $T$ independent draws from the marginal posterior distribution of $L_{jk}$.

A likelihood ratio $L_{jk}$ of 9, with equal prior probabilities of 0.5 on each model, would give a posterior probability of 0.9 for model $j$, which would generally be regarded as quite strong evidence for model $j$ over model $k$. Over the $T$ draws $L_{jk}^{[t]}$, we compute the (empirical) probability that $L_{jk} > 9$; if this is 0.9 or more, we have a high posterior probability of quite strong evidence in favor of model $j$ over model $k$.

It is easier to compute and interpret *deviances* (unfortunately easily confused with frequentist deviances). The deviance for model $M_j$ is $D_j = -2\log L_j(\theta_j)$.

For the $t$th draw $\theta_j^{[t]}$ from the posterior of $\theta_j$, we obtain the $t$th draw of the deviance for model $M_j$:

$$D_j^{[t]} = -2\log L_j(\theta_j^{[t]}).$$

A likelihood ratio $L_{jk}$ of 9 is equivalent to a deviance difference $D_{jk} = D_j - D_k$ of $-2\log 9 = -4.4$. If the empirical $\Pr[D_{jk} < -4.4|\text{data}] > 0.9$, we have a high posterior probability of quite strong evidence for $M_j$ against $M_k$. We extend this approach to model averaging in Section 5.

A formal analysis follows Spiegelhalter et al. (2002). For large samples from regular models with flat priors on the parameters $\theta$, a Taylor expansion of $D(\theta)$ about $\hat{\theta}$ shows that the deviance can be expressed [equation (18) of Spiegelhalter et al. (2002)] as

$$D(\theta) \simeq D(\hat{\theta}) + \chi_s^2,$$

where $s$ is the dimensionality of $\theta$, and $D(\hat{\theta})$ is the frequentist deviance. (For nonnested model comparisons, all integrating constants must be included in the model likelihood.) If the competing deviance distributions are plotted as cdfs on the deviance scale, those with more parameters will have lower slopes because of the increasing variance of the $\chi_s^2$ distribution with $s$.

For comparison of nonnested models 1 and 2, the deviance difference

$$D_{12} = D_1(\theta_1) - D_2(\theta_2) \simeq D_1(\hat{\theta}_1) - D_2(\hat{\theta}_2) + \chi_{s_1}^2 - \chi_{s_2}^2,$$

where the $\chi^2$ variables are independent. Thus, the distribution of the deviance difference $D_{12}$ will be a difference of independent $\chi^2$ variables location-shifted by the frequentist deviance difference $FD_{12}$. The distribution of a difference between independent $\chi^2$ variables does not have a closed-form density, but is very easily simulated.

If both $s_1$ and $s_2$ are large,

$$D_{12} \simeq s_1 - s_2 + \sqrt{2s_1}N(0,1) - \sqrt{2s_2}N(0,1) + FD_{12},$$

that is,

$$D_{12} \simeq s_1 - s_2 + N(0, 2s_1 + 2s_2) + FD_{12}.$$

Here the frequentist deviance difference is penalized by the difference in degrees of freedom, but the *uncertainty* in the comparison—which increases with increasing numbers of parameters—provides a probabilistic measure of certainty about the superiority of one model over the other. This is different from the usual penalized LRT approaches which provide only single-number *decision* criteria without a probabilistic interpretation.

The likelihoods being compared are the *observed-data likelihoods*, obtained by integrating out the random effects. Their computation will in general require numerical integration or MCMC to obtain the parameter posteriors. For the lung cancer data there are no covariates, so a direct integration approach is straightforward. Since there are no individual-level variables, each city population is effectively homogeneous, and the two-level variance component model reduces to an *overdispersion* model in the proportions $p_i$. We first adapt the original analysis by Tsutakawa.

4.1. *Normal logit analysis.* We depart slightly from Tsutakawa's analysis by using the binomial distribution for the number of deaths in each city rather than the Poisson distribution; since the observed rates are very small, and the city populations 100 or more, this makes a negligible difference, as we use the same upper-level normal logit model as Tsutakawa.

The number of deaths $r_i$ is modeled by a binomial random variable $R_i$ with true death proportion $p_i$:

$$R_i \mid p_i, n_i \sim b(n_i, p_i).$$

At the upper level, the logits of the city proportions are modeled by a non-conjugate normal distribution:

$$\theta_i = \log\left[\frac{p_i}{1-p_i}\right] \sim N(\mu, \sigma^2).$$

To compute the likelihood $L_1(\mu, \sigma)$, we integrate over the unobserved $\theta_i$ using a $(K=)20$-point Gaussian quadrature, with masses $\pi_k$ at mass-points $z_k$, $k = 1, \ldots, K$:

$$L_1(\mu, \sigma) = \prod_{i=1}^{N} \binom{n_i}{r_i} \int_{-\infty}^{\infty} \frac{e^{\theta_i r_i}}{(1+e^{\theta_i})^{n_i}} \frac{1}{\sqrt{2\pi}\sigma} \exp\left\{-\frac{1}{2}\frac{(\theta_i-\mu)^2}{\sigma^2}\right\} d\theta_i$$

$$= \prod_{i=1}^{N} \binom{n_i}{r_i} \int_{-\infty}^{\infty} \frac{e^{(\mu+\sigma z_i)r_i}}{(1+e^{\mu+\sigma z_i})^{n_i}} \frac{1}{\sqrt{2\pi}} \exp\left\{-\frac{1}{2}z_i^2\right\} dz_i$$

$$\doteq \prod_{i=1}^{N} \binom{n_i}{r_i} \sum_{k=1}^{K} \frac{e^{(\mu+\sigma z_k)r_i}}{(1+e^{\mu+\sigma z_k})^{n_i}} \pi_k.$$

We compute the likelihood numerically over an equally spaced grid of $G = 100 \times 100$ values $(\mu_{[g]}, \sigma_{[g]})$ in the region of appreciable likelihood: $\mu \in (-4.56, -4.20), \sigma \in (0.08, 0.40)$, sum the likelihoods over the grid and normalize to give the posterior *mass* function $\pi(\mu, \sigma | \mathbf{y})$ of $(\mu, \sigma)$ for flat priors on the grid, as shown in Figure 3.

The maximum likelihood estimates of $\mu$ and $\sigma$, and the maximized log-likelihood over the grid are $\hat{\mu} = -4.787, \hat{\sigma} = 0.2368$ and $-181.254$; the frequentist deviance is 362.51. These discrete MLEs are close, but not identical, to those reported by Tsutakawa: $\hat{\mu} = -4.733, \hat{\sigma} = 0.2384$. The figure strongly suggests multi-modality in the distribution, but this causes no difficulty in the subsequent analysis. Careful inspection shows that Tsutakawa's estimates are near a saddle point, and that the likelihood has four local maxima, with values of $(\mu, \sigma)$ and log-likelihoods given by $(-4.787, 0.2368, -181.254)$, $(-4.708, 0.2016, -181.397)$, $(-4.776, 0.3072, -181.425)$, $(-4.668, 0.2432, -181.612)$. The log-likelihood at Tsutakawa's estimates is $-182.059$. These small differences have very little effect on the area posterior densities.

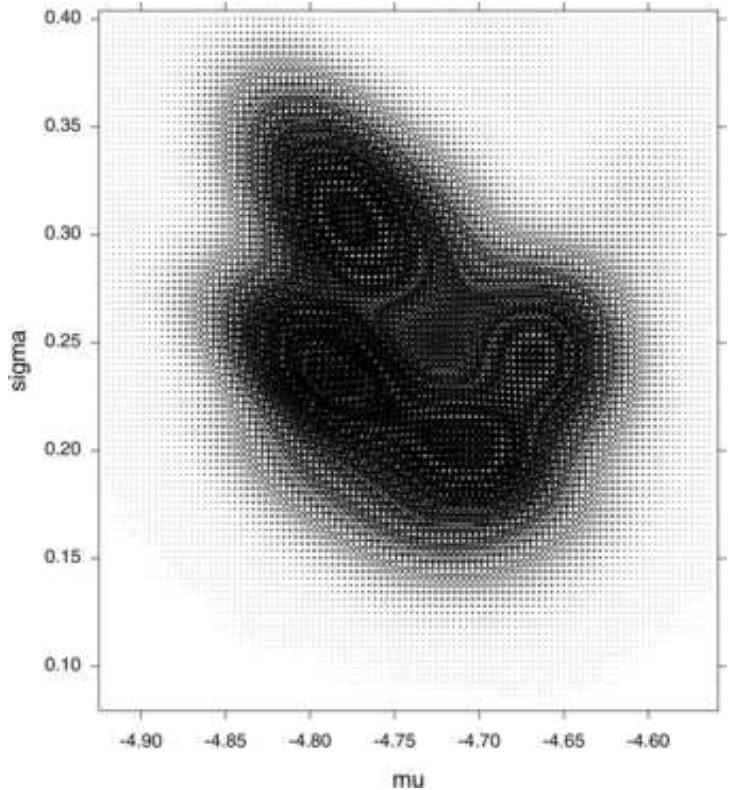

Fig. 3. *Joint posterior mass function of $(\mu, \sigma)$.*

We note for later discussion that, by weighting $-2$ times the log-likelihoods $\ell_1(\mu_{[g]}, \sigma_{[g]})$ at each grid point by $\pi(\mu_{[g]}, \sigma_{[g]} | \mathbf{y})$, their posterior probabilities, and summing, we obtain the *posterior mean of the deviance* used in the DIC [Spiegelhalter et al. (2002)]. For the normal model the posterior mean deviance is 364.32. This may be combined with the maximized log-likelihood to give one version of the *effective number of parameters* $p_D$ defined by equation (10) of Spiegelhalter et al. (2002):

$$p_D = \overline{D(\theta)} - D(\tilde{\theta});$$

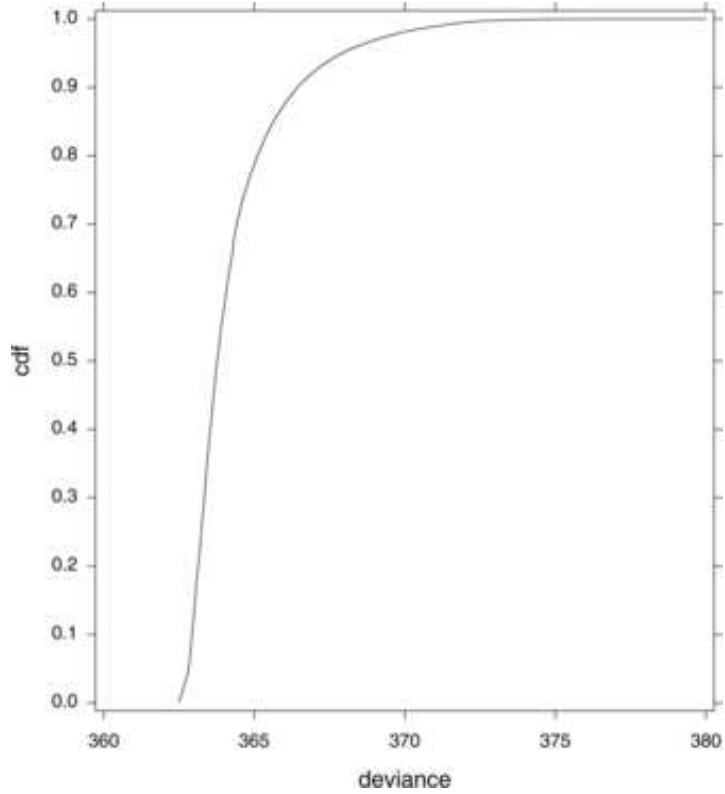

Fig. 4. *Deviance distribution for normal model.*

we use for the second term the deviance at the MLEs rather than at the posterior mean, since the latter is not a grid point in general and would require additional computation. Here the second term is 362.51, giving $p_D = 1.81$, and $DIC_1 = 366.13$.

However, Figure 3 provides much more information, since it gives the *exact* (up to the grid resolution) posterior distribution of the deviance, without any Monte Carlo simulation. Sorting the deviances

$$D_2(\mu_{[g]}, \sigma_{[g]}) = -2\ell_2(\mu_{[g]}, \sigma_{[g]})$$

into increasing order with their corresponding posterior probabilities and cumulating the latter, we obtain the cdf of the deviance, as shown in Figure 4.

Conditional on the $i$th area data $r_i, n_i$ and the parameters $\mu, \sigma$, the posterior density of $p_i$ has the form

$$\pi(\theta_i|\mu,\sigma,r_i) = c(\mu,\sigma) \binom{n_i}{r_i} p_i^{r_i}(1-p_i)^{n_i-r_i} \exp\left\{-\frac{1}{2}\frac{(\theta_i-\mu)^2}{\sigma^2}\right\}$$

$$= c(\mu,\sigma) \binom{n_i}{r_i} \frac{e^{r_i\theta_i}}{(1+e^{\theta_i})^{n_i}} \exp\left\{-\frac{1}{2}\frac{(\theta_i-\mu)^2}{\sigma^2}\right\}.$$

Expanding the log density in $\theta_i$ about $\hat{\theta}_i = \log[r_i/(n_i - r_i)]$ up to second-order terms, it is easily shown that, to this order of approximation,

$$\theta_i \mid r_i, \mu, \sigma \sim N\left(\frac{\psi_i\hat{\theta}_i + \psi\mu}{\psi_i + \psi}, \frac{1}{\psi_i + \psi}\right),$$

where $\psi_i$ and $\psi$ are the sample and prior precisions of $\theta_i$:

$$\psi_i = n_i\hat{p}_i(1-\hat{p}_i), \qquad \psi = 1/\sigma^2.$$

This approximation fails if $r_i = 0$; in this case we set $r_i = 0.5$. The level of approximation is not affected by the distribution of $(\mu, \sigma)$.

The posterior densities for the area random effects $p_i$ are most easily computed by Gaussian kernel smoothing of $T = 10{,}000$ random values $p_i^{[t]}$, generated from the normal distributions obtained by substituting $T$ random draws $(\mu^{[t]}, \sigma^{[t]})$ into the area posterior distributions:

$$p_i^{[t]} = \frac{\psi_i\hat{\theta}_i + \psi^{[t]}\mu^{[t]}}{\psi_i + \psi^{[t]}} + z^{[t]}/\sqrt{\psi_i + \psi^{[t]}},$$

where $\psi^{[t]} = 1/\sigma^{[t]2}$ and $z^{[t]}$ is a random draw from $N(0,1)$.

Figure 5 shows the results for the five cities above for $T = 10{,}000$.

Comparison with Tsutakawa's Figure 1 shows close agreement for all five cities, though the vertical scales are different.

We now re-analyze the data with a beta distribution for the area proportions, which gives simply computed forms for the likelihood and the posterior distributions.

4.2. *Beta-binomial analysis.* At the upper level, the city proportions are modeled by a conjugate beta distribution:

$$P_i \mid a, b \sim Beta(a, b),$$

with density function

$$f(p) = p^{a-1}(1-p)^{b-1}/B(a,b), \qquad a, b > 0,$$

where $B(a,b)$ is the complete beta function

$$B(a,b) = \Gamma(a)\Gamma(b)/\Gamma(a+b).$$

The beta-binomial likelihood is denoted by

$$L_2(a,b) = \prod_{i=1}^m \left[\binom{n_i}{r_i} B(r_i + a, n_i - r_i + b)/B(a,b)\right].$$

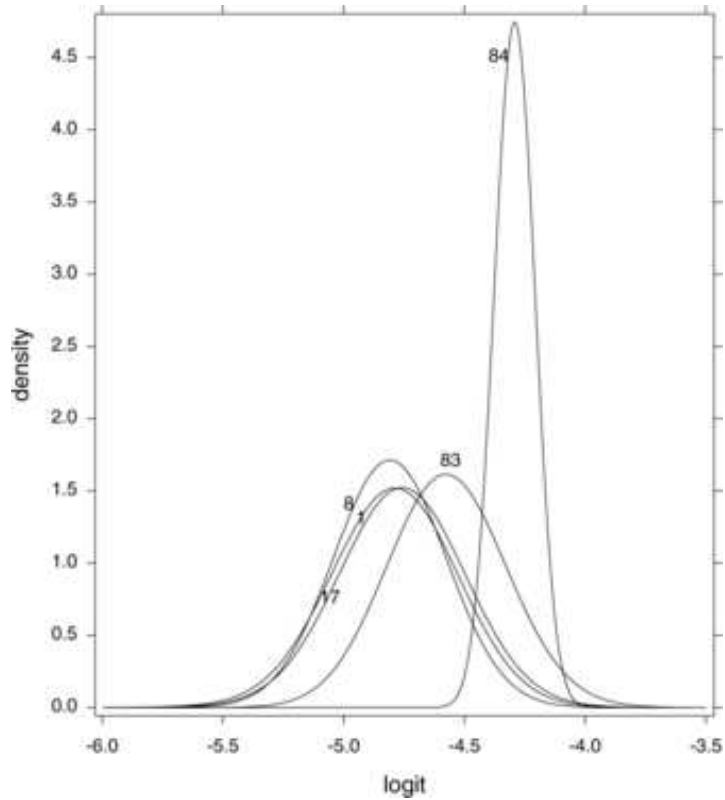

FIG. 5. *Normal model posterior densities for five cities.*

The likelihood in this parametrization is very highly correlated in $(a, b)$; we use instead the (mean, standard deviation) parametrization, with

$$\mu_\beta = a/(a+b), \qquad \sigma_\beta = \sqrt{ab/[(a+b)^2(a+b+1)]}.$$

The likelihood, shown in Figure 6, is nearly orthogonal in these parameters. We use this form of the likelihood for subsequent computation.

We compute the likelihood numerically over a grid of $G = 100 \times 100$ values $(\mu_{\beta[g]}, \sigma_{\beta[g]})$ in the region of appreciable likelihood: $\mu_\beta \in (0.007, 0.011), \sigma_\beta \in (0.001, 0.004)$, sum the likelihoods over the grid and normalize to give the posterior mass function $\pi(\mu_\beta, \sigma_\beta | \mathbf{y})$ of $(\mu_\beta, \sigma_\beta)$ on the grid. The maximized log-likelihood is $-181.486$, with frequentist deviance $362.97$.

As for the normal model, by weighting the log-likelihoods at each grid point by $-2$ times their posterior probabilities and summing, we obtain the posterior mean of the deviance for the beta model, which is $364.99$, slightly larger than that for the normal model, of $364.32$. The effective number of parameters is $p_D = 2.02$, and $DIC_2 = 367.01$.

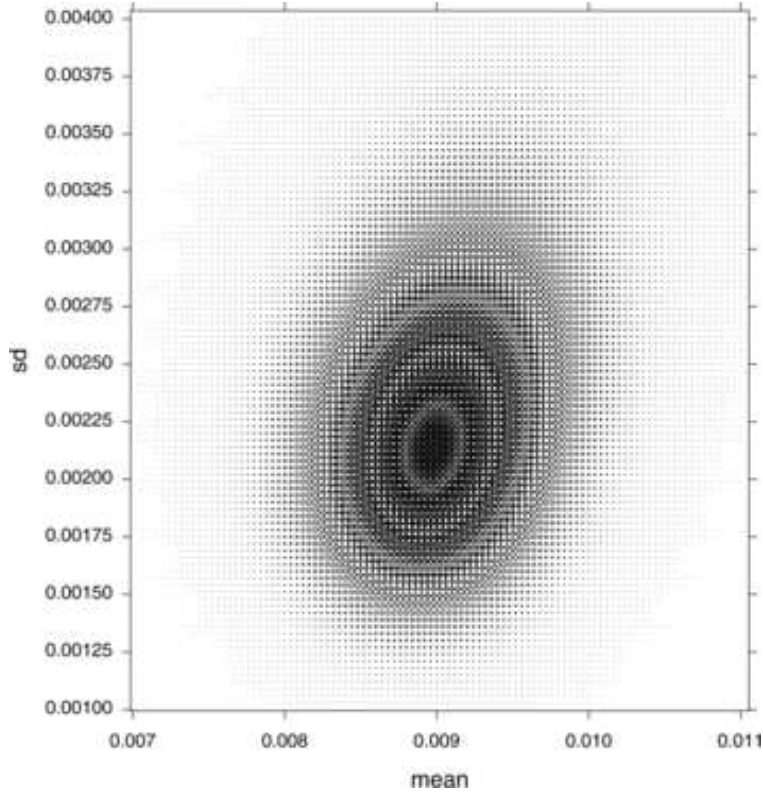

Fig. 6. *Joint posterior mass function of beta mean and SD.*

The full posterior distribution of the beta deviance $D_2(\mu_\beta, \sigma_\beta) = -2 \log L_2(\mu_\beta, \sigma_\beta)$, computed as for the normal model, is shown (dotted curve) in Figure 7, together with that of the normal model 1 (solid curve). We discuss the comparison of these models in the next section.

Conditional on the $i$th area data $r_i, n_i$ and the parameters $a, b$, the posterior distribution of $p_i$ is again beta:

$$\pi(p_i \mid a, b, r_i) = p_i^{r_i+a-1}(1-p_i)^{n_i-r_i+b-1}/B(r_i + a, n_i - r_i + b).$$

The posterior densities for the area random effects $p_i$ are again computed by Gaussian kernel smoothing of $T = 10{,}000$ random values $p_i^{[t]}$, generated from $T$ random draws $(\mu_\beta^{[t]}, \sigma_\beta^{[t]})$ from their posterior distribution which are converted to $T$ random values $(a^{[t]}, b^{[t]})$ of $(a, b)$. We finally draw the $T$ random values $p_i^{[t]}$ of $p_i$, one each from the $T$ beta distributions with indices $r_i + a^{[t]}, n_i - r_i + b^{[t]}$ for the given $i$.

The $T$ values of $p_i$ are transformed to the logit scale for ease of inspection and consistency with Tsutakawa's analysis; posterior densities for individual cities are then computed using Gaussian kernel densities with bandwidths chosen to give smooth densities. Figure 8 shows the five beta posteriors (dotted curves) together with the normal posteriors from Figure 6 (solid curves). The city numbers are placed at the intersection of the two densities.

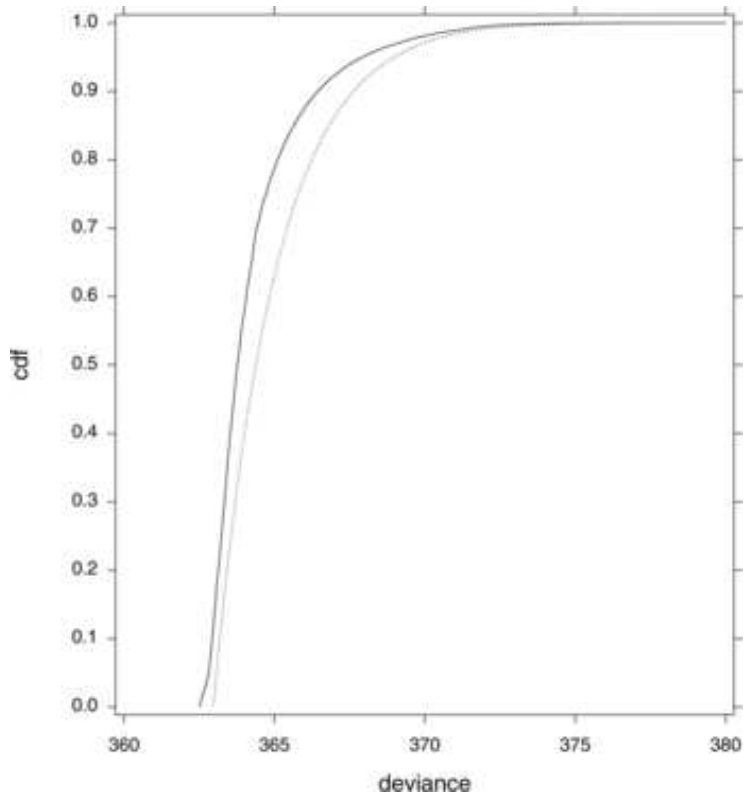

Fig. 7. *Beta (dotted) and normal (solid) deviances.*

The beta posteriors are slightly less concentrated than the normal posteriors except for city 84, and show slightly more shrinkage toward the mean.

Since the posterior conclusions from the beta distribution differ somewhat from those from the normal, we need to decide whether the data support one model over the other.

**5. Model comparisons via deviances.** For the comparison of these models, and of other models for the true proportions, we compare the model deviances $D = -2 \log L$.

The two models for the $p_i$ considered above are not the only possible models. Following Spiegelhalter et al. (2002), we consider four possible models for the $p_i$:

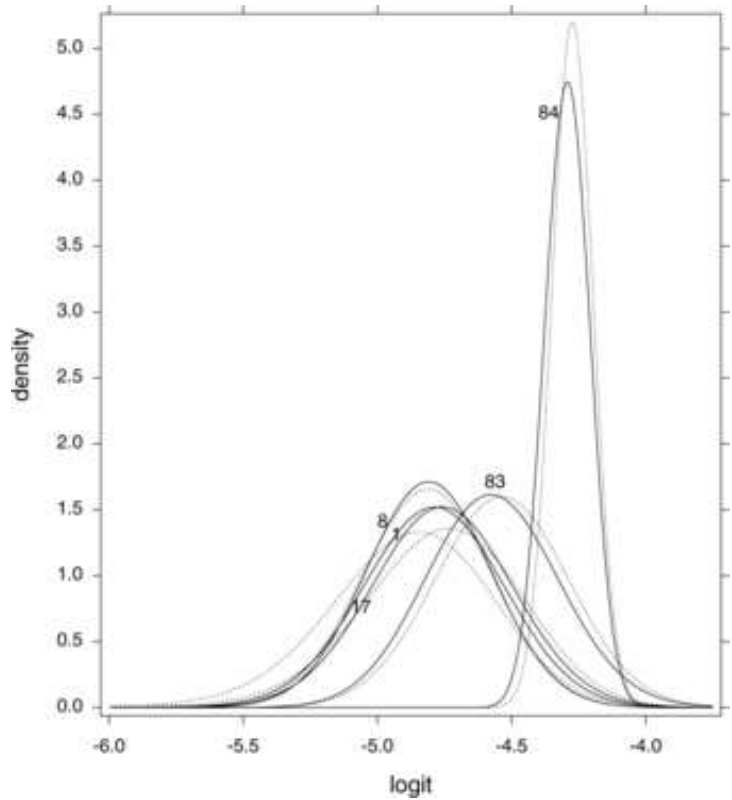

Fig. 8. *Beta (dotted) and normal (solid) posterior densities for five cities.*

- Model 1—the *normal logit* model: $\text{logit}\, p_i \sim N(\mu, \sigma^2)$;
- Model 2—the *beta* model: $p_i \sim B(a, b)$;
- Model 3—the *null* model: $p_i \equiv p$;
- Model 4—the *saturated* model: $p_i$ all different and unrelated.

Models 3 and 4 are different from Models 1 and 2 in that there is no actual parametric model for the variation of the $p_i$ across the cities—each city has its own single parameter under Model 4, and all cities have the same single parameter under Model 3.

We compare directly the deviance distributions under each model. We first apply this approach to the normal and beta deviance distributions; we show all distributions in Figure 10. The beta deviance has a lower slope and is consistently to the right of the normal deviance, so the normal model is preferred, but not very strongly. To compare the models directly we draw 10,000 random values $D_1^{[t]}, D_2^{[t]}$ from each deviance distribution and compute the deviance differences $D_{12}^{[t]} = D_2^{[t]} - D_1^{[t]}$ (beta − normal) from the 10,000 (unordered) values for each model. The cdf of the deviance difference distribution is shown in Figure 9.

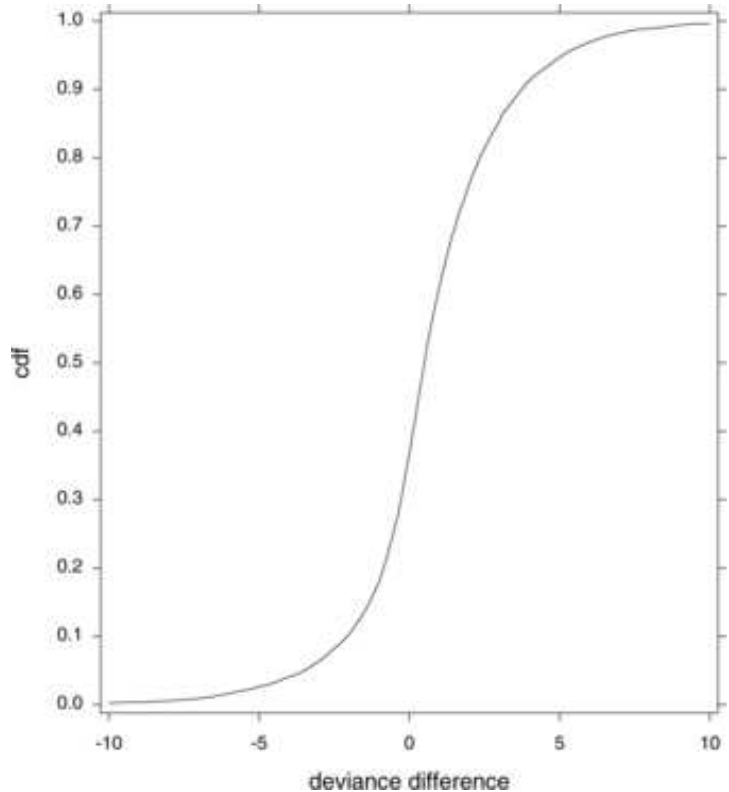

Fig. 9. $Beta - normal\ deviance$.

The distribution has its median at 0.505, and the 95% credible interval for the true difference, computed from the 250th and 9750th ordered differences [Congdon (2005)] is $(-5.125, 6.378)$. The estimated probability that the normal deviance is smaller than the beta is 0.6332: we cannot confidently choose between these models.

The deviance constructions for Models 3 and 4 are different from those for Models 1 and 2. The null Model 3 likelihood given $p$ is

$$L_3(p) = \prod_{i=1}^{m} \left[ \binom{n_i}{r_i} p^{r_i}(1-p)^{n_i-r_i} \right]$$
$$= \left[ \prod_{i=1}^{m} \binom{n_i}{r_i} \right] p^R (1-p)^{N-R},$$

where $R = \sum_i r_i, N = \sum_i n_i$.

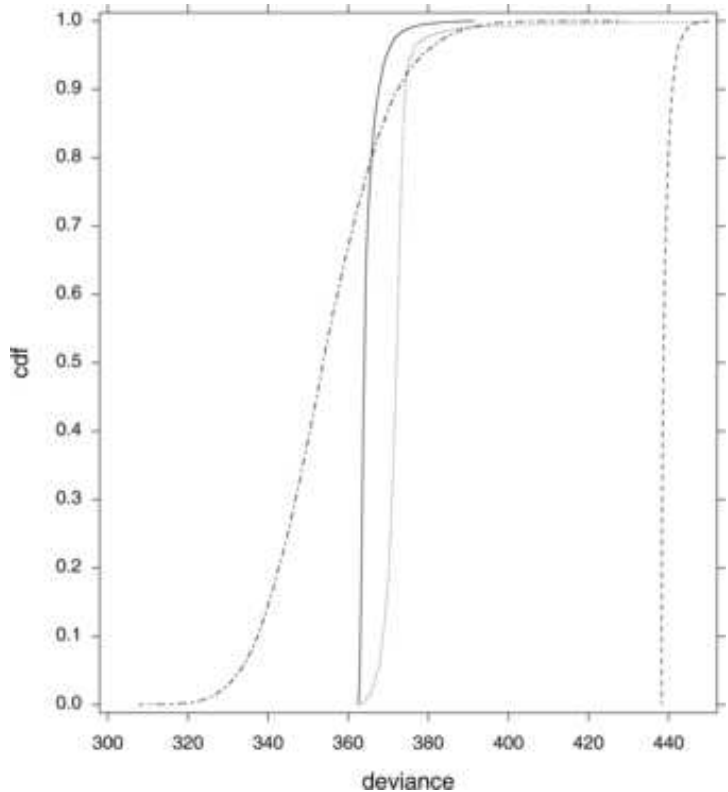

Fig. 10. *Deviances for null (dashed), normal (solid), beta (dotted) and saturated (dot–dashed) models.*

For this one-parameter model, which is effectively a single-sample model, the prior represents uncertainty about the value of $p$, not the variability of different $p_i$ across cities. We use a uniform prior for $p$, giving a $Beta(R + 1, N - R + 1)$ posterior distribution of $p$. We draw 10,000 random values $p^{[t]}$ from this posterior, and for each compute the likelihood $L_3(p^{[t]})$ and the corresponding deviance $D_3^{[t]}$.

For the saturated Model 4, the prior specification is similar. Each city has its own $p_i$, unrelated to the others, so the uniform prior is used for each city independently, giving the set of $Beta(r_i + 1, n_i - r_i + 1)$ posterior distributions of $p_i$. We draw 10,000 random values $p_i^{[t]}$ from each posterior, and for each $m$ and each $i$ compute the likelihood contribution $L_{4i}(p_i^{[t]})$ and the corresponding deviance $D_4^{[t]} = -2\sum_i \log L_{4i}(p_i^{[t]})$.

The cdfs of all four deviance distributions are shown in Figure 10.

Model 3 is nearly 40 deviance units to the right of the normal and beta—it is immediately clear that the null model is untenable. We exclude it from further consideration. The deviance distribution for the saturated model crosses those for the normal and beta models, indicating no strong preference for the saturated model over the parametric models. For the better-fitting normal model, the distribution of the deviance difference (normal deviance − saturated deviance) is shown in Figure 11.

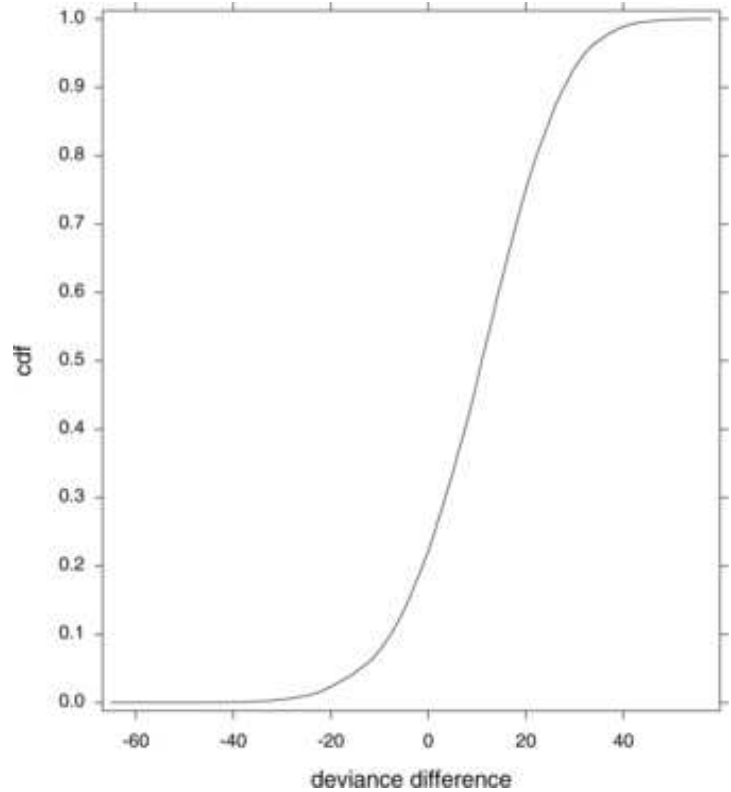

Fig. 11. *Deviance difference normal − saturated model.*

Of the 10,000 differences, 2216 are positive, so the estimated probability that the saturated deviance is smaller is 0.7784. This is not compelling evidence against the normal model; since the saturated model allows no "borrowing of strength" across cities, we retain both the normal and beta models as candidates for the upper-level distribution, in addition to the saturated model.

**6. Posterior model averaging.** *Model averaging* is frequently proposed [see, e.g., Hoeting et al. (1999)] for posterior inference about a common parameter (like the mean) across several competing models. The posterior densities under each model are averaged with respect to the posterior prob-abilities of each model, based on their marginal likelihoods integrated over the prior distributions of the parameters.

This process uses the integrated likelihoods; as with model choice, we use the posterior *distributions* of the likelihoods to provide an averaged posterior density, but with respect to posterior probabilities of each model which are themselves random variables rather than fixed constants. The model comparison approach of Section 3 is simply extended to deal with this.

As before, $\theta_j$ are the model parameters for model $M_j$, $\theta_j^{[t]}$ are the $T$ draws from the posterior distribution of $\theta_j$ given the data $\mathbf{y}$, $L_j^{[t]}$ are the corresponding $T$ draws from the posterior distribution of $L_j$, and $p_{ij}^{[t]}$ are the corresponding $T$ values of the random effect $p_i$ for area $i$ under $M_j$.

Let $\pi_j$ be the prior probability of Model $j$; the $t$th draw $\pi^{[t]}(M_j|\mathbf{y})$ from the posterior probability of Model $j$ is then

$$\pi^{[t]}(M_j|\mathbf{y}) = \pi_j L_j^{[t]} \Big/ \sum_k \pi_k L_k^{[t]}.$$

The averaged density $p_{\mathrm{ave},i}$ for $p_i$ is then defined through the $T$ simulated values $p_{\mathrm{ave},i}^{[t]}$ produced by the algorithm: for each $i$ and each $m$:

1. compute $\pi^{[t]}(M_j|\mathbf{y})$;
2. with probability $\pi^{[t]}(M_j|\mathbf{y})$, set $p_{\mathrm{ave},i}^{[t]} = p_{ij}^{[t]}$.

This looks complicated notationally but is quite simple: the likelihoods for each draw define the posterior probabilities for each model for this draw, and we simply choose the model $j$ random effect draw with the model $j$ posterior probability draw [Congdon (2005, 2006)].

In the computation of the averaged density we exclude the null model, and include the other three. Prior probabilities for the three distributions are taken as equal, though as described above it is a simple matter to change them. We show below several graphs of the city posterior rate densities on the logit scale for the five cities shown earlier. Figure 12 shows the averaged densities, Figure 13 shows the averaged (solid) and local area (dot-dashed) densities from Figure 2, and Figure 14 shows the averaged (solid) and normal (dot-dashed) densities.

The sharpness of the normal area posteriors is damped by the averaging process, because in the 10,000 draws from each posterior deviance distribution, the saturated model likelihood is generally superior to the normal or beta likelihood. For the large city 84 there is little difference in the posteriors because of the very large city size; this city "lends strength" to the small cities in the normal model which are substantially shrunk toward the posterior mean. The generally superior likelihood for the saturated model shows that neither the normal nor the beta provides a convincing representation of the random effect distribution, and so conclusions about individual random effects need to be based substantially on the local area rate.

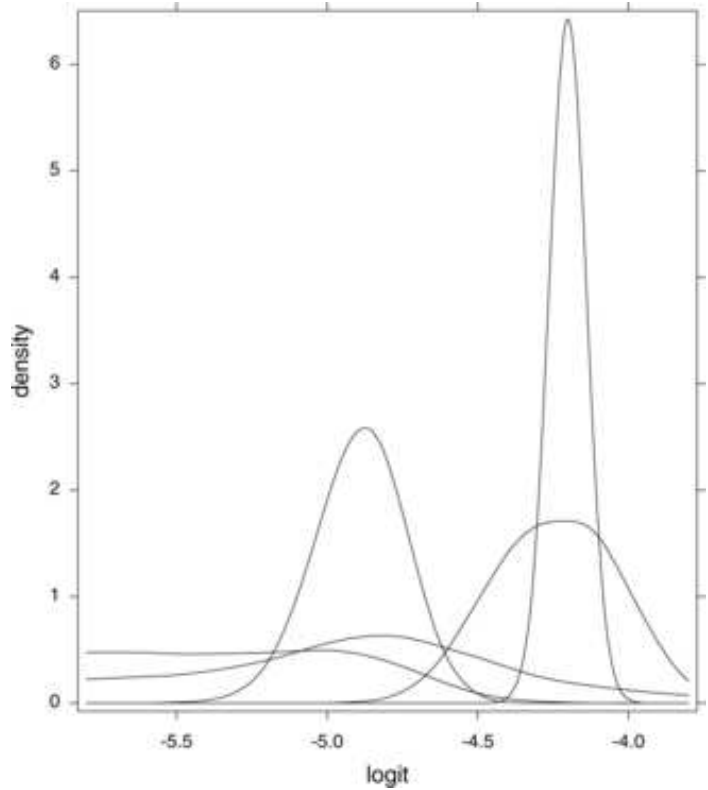

Fig. 12. *Averaged posteriors, five cities.*

**7. Model averaging for a "typical" city.** The analysis above assumes that our inferential interest is in the individual city rates. This is the most common use of the two-level model. If, however, (or in addition) we wish to draw conclusions about a "typical" city, this is done using the among-city model—either normal or beta. The saturated model does not provide information about a "typical" city, because the city rates are unrelated under this model.

The definition of a "typical" city is unclear; we take for illustration a *random* city, in the sense of a random draw from the among-city model. Under the normal model, the random logit rate will be $\mu + \sigma Z$, where $Z \sim N(0, 1)$, while under the beta model for $p$, it will be a random draw from

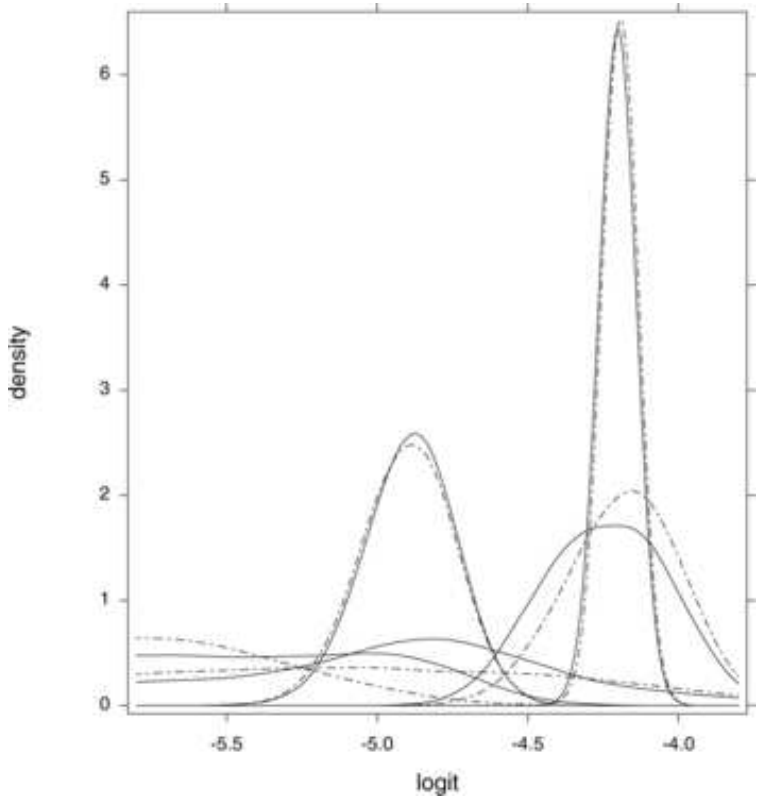

Fig. 13. *Averaged (solid) and local (dot-dashed) posteriors, five cities.*

the transformed beta density

$$f(\theta) = \frac{e^{\theta a}}{(1+e^\theta)^{a+b}}.$$

So the posterior distribution of the random city rate under the normal model is given by the empirical distribution of the $T$ random draws $\mu^{[t]} + \sigma^{[t]} Z^{[t]}$, while that for the beta model is that of the $T$ random draws $\theta^{[t]}$ from $Beta(a^{[t]}, b^{[t]})$. To model average these, we draw the normal $\mu^{[t]} + \sigma^{[t]} Z^{[t]}$ with posterior probability $\pi^{[t]}(N|\mathbf{y})$, and the beta $\theta^{[t]}$ with posterior probability $1 - \pi^{[t]}(N|\mathbf{y})$, where only these two models have positive posterior probability.

**8. Discussion.** These conclusions may seem surprising. The idea of "borrowing strength" is well established in the Bayesian and non-Bayesian literature for random effect models. The difficulty with the lung cancer data is that the "strong" cities from which strength may be borrowed have essentially only two support points, since cities 4 and 44 have observed rates (0.00742 and 0.00867) which are very similar and near the median observed rate (though nearly the smallest of the normal posterior mean rates), while city 84 has almost the highest observed rate (0.01484), and has the highest normal posterior mean rate.

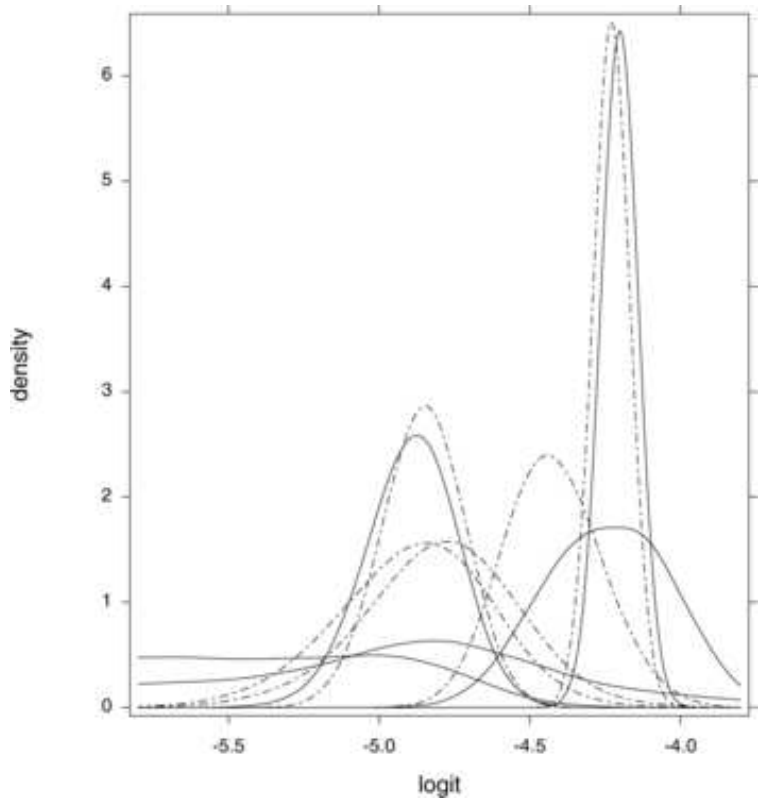

Fig. 14. *Averaged (solid) and normal (dot-dashed) posteriors, five cities.*

The small cities with limited data cannot shrink effectively toward either support point, since their rates are not well identified with these points, and the generally higher likelihood for the saturated model accentuates this effect. The model averaged rate distributions remain diffuse, though less so than the single city rates.

What has been missing in the use of these models is a direct and straightforward assessment of the appropriateness, or goodness of fit, of the upper-level model. This has been done in Bayesian analysis mostly through Bayes factors, with their attendant prior sensitivity difficulties, or through the DIC, with its definitional "focus" difficulties and ambiguous penalty.

The comparison of models through their posterior likelihood or deviance distributions provides a straightforward way, not only to compare different parametric models, but also to evaluate the quality of reproduction of the local area rates by the parametric models.

It is not, however, necessary to "pick the best model": the posterior likelihood model comparison extends directly to Bayesian model averaging, and provides a compromise among the well-supported competing models. In this approach we treat inference about the "typical" city rate and about the individual city rates in the same way, though the relevant distributions are marginal instead of conditional.

**9. Conclusion.** The comparison of deviance distributions allows the expression of model comparisons through likelihood ratios, as for simple null and alternative hypotheses, while allowing automatically for the complexity of the model: unnecessarily complex models have more diffuse likelihoods and so the approach of explicit penalization of the maximized likelihood through some function of the number of model parameters is not necessary. Incorporation of informative parameter or model priors is simple and straightforward. A further example of deviance distribution comparisons on the galaxy data of Roeder (1990) is given in the discussion of Ridall et al. [(2007), pages 264–265] by Aitkin.

The choice of a suitable upper-level model for this example is informed by Tsutakawa's original analysis of the data; for this simple data set the beta distribution provides an alternative model which performs nearly as well as the normal model, but both these models are (nonsignificantly) inferior to the saturated model, despite the diffuseness of its deviance distribution from the 84 cities.

This analysis is simply adapted to provide model-averaged posterior shrinkage when several upper-level models are well supported by the data; these model-averaged posteriors provide a compromise reflecting the uncertainty about the appropriate model.

**Acknowledgments.** We appreciate initial support for this work from the National Center for Education Statistics, encouragement from Gary Phillips, the then Acting Commisioner of NCES, and helpful comments from Steve Fienberg, Robert Wolpert, Kerrie Mengersen, Ross McVinish and an Associate Editor and referee.

## REFERENCES

AITKIN, M. (1997). The calibration of $P$-values, posterior Bayes factors and the AIC from the posterior distribution of the likelihood (with discussion). *Statist. Comput.* **7** 253–272.

AITKIN, M. (1999). A general maximum likelihood analysis of variance components in generalized linear models. *Biometrics* **55** 117–128. MR1705676

# BAYESIAN MODEL COMPARISON 23

AITKIN, M., BOYS, R. J. and CHADWICK, T. (2005). Bayesian point null hypothesis testing via the posterior likelihood ratio. *Statist. Comput.* **15** 217–230. MR2147554

CARLIN, B. P. and LOUIS, T. A. (1996). *Bayes and Empirical Bayes Methods for Data Analysis*. Chapman and Hall, London. MR1427749

CELEUX, G., FORBES, F., ROBERT, C. P. and TITTERINGTON, D. M. (2006). Deviance information criteria for missing data models (with discussion). *Bayesian Anal.* **1** 651–706. MR2282197

CONGDON, P. (2005). Bayesian predictive model comparison via parallel sampling. *Comput. Statist. Data Anal.* **48** 735–753. MR2133574

CONGDON, P. (2006). Bayesian model comparison via parallel model output. *J. Statist. Comput. Simul.* **76** 149–165. MR2224357

DEMPSTER, A. P. (1974). The direct use of likelihood in significance testing. In *Proc. Conf. Foundational Questions in Statistical Inference* (O. Barndorff-Nielsen, P. Blaesild and G. Sihon, eds.) 335–352. Kluwer, Hingham, MA. MR0408052

DEMPSTER, A. P. (1997). The direct use of likelihood in significance testing. *Statist. Comput.* **7** 247–252.

FOX, J.-P. (2005). Multilevel IRT using dichotomous and polytomous response data. *Brit. J. Math. Statist. Psych.* **58** 145–172. MR2196136

HOETING, J. A., MADIGAN, D., RAFTERY, A. and VOLINSKY, C. T. (1999). Bayesian model averaging: A tutorial. *Statist. Sci.* **14** 382–417. MR1765176

RIDALL, P. G., PETTITT, A. N., FRIEL, N., HENDERSON, R. and MCCOMBE, P. (2007). Motor unit number estimation using reversible jump Markov chain Monte Carlo methods (with discussion). *J. Roy. Statist. Soc. Ser. C* **56** 235–269. MR2370990

ROEDER, K. (1990). Density estimation with confidence sets exemplified by superclusters and voids in the galaxies. *J. Amer. Statist. Assoc.* **85** 617–624.

SPIEGELHALTER, D. J., BEST, N. G., CARLIN, B. P. and VAN DER LINDE, A. (2002). Bayesian measures of model complexity and fit (with discussion). *J. Roy. Statist. Soc. Ser. B* **64** 583–639. MR1979380

TREVISANI, M. and GELFAND, A. E. (2003). Inequalities between expected marginal log likelihoods with implications for likelihood-based model comparison. *Canadian J. Statist.* **31** 239–250. MR2030122

TSUTAKAWA, R. K. (1985). Estimation of cancer mortality rates: A Bayesian analysis of small frequencies. *Biometrics* **41** 69–79. MR0793434

M. AITKIN
C. C. LIU
SCHOOL OF BEHAVIOURAL SCIENCE
UNIVERSITY OF MELBOURNE
AUSTRALIA
E-MAIL: murray.aitkin@ms.unimelb.edu.au
     charles.liu@muarc.monash.edu.au

T. CHADWICK
INSTITUTE OF HEALTH AND SOCIETY
NEWCASTLE UNIVERSITY
UNITED KINGDOM
E-MAIL: t.j.chadwick@newcastle.ac.uk